\documentclass{elsart}

\usepackage{graphics}
\usepackage{psfig}
\usepackage{amsmath} 
\usepackage{amssymb}
\usepackage{enumerate}
\usepackage[dvips]{graphicx}

\renewcommand{\thefootnote}{\arabic{footnote}}

\newcommand{\gae}{($\gamma$,$\eta$) }

\newcommand{\gpep}{$\gamma p\rightarrow\eta p$}
\newcommand{\gnen}{$\gamma n\rightarrow\eta n$}

\newcommand{\siin}{$S_{11}$(1535)}
\newcommand{\ssiin}{$S_{11}$(1650)}

\newcommand{\dn}{$D_{15}$(1675)}
\newcommand{\pn}{$P_{11}$(1710)}
\newcommand{\sii}{$S_{11}$}
\newcommand{\Eg}{$E_\gamma$ }

\begin{document}
\begin{frontmatter}
\title{Photoproduction of $\eta$-mesons off C and Cu nuclei 
for photon energies below 1.1~GeV}

\author[LNS]{T.~Kinoshita},
\ead{tadashi@lns.tohoku.ac.jp}
\author[LNS]{H.~Yamazaki},
\author[LNS]{H.~Fukasawa},
\author[LNS]{K.~Hirota\thanksref{now-RIKEN}},
\author[LNS]{T.~Ishikawa},
\author[LNS]{J.~Kasagi},
\author[LNS]{A.~Kato\thanksref{now-Donen}},
\author[LNS]{T.~Katsuyama},
\author[LNS]{K.~Kino\thanksref{now-CNS}},
\author[LNS]{F.~Miyahara},
\author[LNS]{T.~Nakabayashi},
\author[LNS]{K.~Nawa},
\author[LNS]{K.~Okamura},
\author[LNS]{Y.~Saitoh},
\author[LNS]{K.~Satou},
\author[LNS]{M.~Sengoku},
\author[LNS]{H.~Shimizu},
\author[LNS]{K.~Suzuki},
\author[LNS]{S.~Suzuki},
\author[LNS]{T.~Terasawa},
\author[Tohoku]{H.~Kanda},
\author[Tohoku]{K.~Maeda},
\author[Tohoku]{T.~Takahashi\thanksref{now-KEK}},
\author[Yama]{Y.~Aruga},
\author[Yama]{T.~Fujinoya},
\author[Yama]{A.~Iijima},
\author[Yama]{M.~Itaya},
\author[Yama]{Y.~Ito},
\author[Yama]{T.~Iwata},
\author[Yama]{H.~Kato},
\author[Yama]{T.~Kawamura},
\author[Yama]{T.~Michigami},
\author[Yama]{M.~Moriya},
\author[Yama]{T.~Sasaki},
\author[Yama]{Y.~Tajima},
\author[Yama]{S.~Takita},
\author[Yama]{T.~Noma},
\author[Yama]{M.~Yamamoto},
\author[Yama]{H.Y.~Yoshida},
\author[Yama]{Y.~Yoshida},
\author[Ichi]{O.~Konno},
\author[Nichi]{T.~Maruyama},
\author[JASRI]{T.~Yorita}
\address[LNS]{\it
Laboratory of Nuclear Science, Tohoku University,  Sendai~982-0826,
Japan}
\address[Tohoku]{\it
Department of Physics, Tohoku University, Sendai~980-8578, Japan
}
\address[Yama]{\it
Department of Physics, Yamagata University, Yamagata~990-8560, Japan
}
\address[Ichi]{\it
Ichinoseki National College of Technology, Ichinoseki~021-8511, Japan
}
\address[Nichi]{\it
College of Bioresource Sciences, Nihon University,
Fujisawa~252-8510, Japan
}
\address[JASRI]{\it
Japan Synchrotoron Radiation Research Institute, Mikazuki~679-5198, Japan
}
\thanks[now-RIKEN]{
Present address: RIKEN, Wako~351-0198, Japan 
}
\thanks[now-Donen]{
Present address: Japan Cycle Development Institute,
Tokaimura~319-1194, Japan
}
\thanks[now-CNS]{
Present address: 
Center for Nuclear Study (CNS), University of Tokyo, Wako~351-0198, Japan
}
\thanks[now-KEK]{
Present address: High Energy Accelerator Research Organization (KEK),
Tsukuba~305-0801, Japan
}

\begin{abstract}
The $\eta$-meson photoproduction cross sections 
have been measured on the C and Cu  
targets for the photon energies between 600 and 1100~MeV 
to investigate the behavior of the \siin\  resonance in a nucleus.
The excitation functions of the cross section as well as  
the angular and momentum distributions of $\eta$-mesons are in 
quantitative agreement with the Quantum Molecular Dynamics (QMD) model
 calculations in which the 
$\eta$-meson emission processes other than the \siin\  resonance are also
incorporated as proposed in the $\eta$-MAID model. 
It is shown that the excitation of the \dn\  resonance might 
play an important role for $E_{\gamma}>900$~MeV. 
\end{abstract}

\end{frontmatter}



The behavior of hadrons in the nuclear medium is one of the most
intriguing topics in hadron and nuclear physics.
Photon induced reactions are advantageous to 
producing hadrons 
deeply inside a nucleus 
because photons are hardly absorbed.
Modifications in appearances may always be observed. 
Most of them originate simply from the basic effects of the nuclear
medium, 
such as the Fermi motion of nucleons, 
Pauli blocking of the final state and collisions with nucleons.
In addition, an interesting possibility has been proposed;  
i.e., mass modification arising from 
partial restoration of chiral symmetry
in the nuclear medium~\cite{Hatsuda,Jido}.
The effects of the mass change of the $\rho$-meson have been  
studied in $\rho$ photoproduction 
on nuclear targets~\cite{RhoPhoto,Mainz_D13} 
as well as in the hadron reactions~\cite{RhoCERN}. 
However, mass change of baryons 
has not been studied well 
except for the $\Delta$ resonance~\cite{delta}.

The \siin\  resonance is proposed to be a
candidate of the chiral partner of the ground state nucleon, 
and its resonance energy is expected to shift down
by about 100~MeV in the nuclear medium where chiral symmetry is partially
 restored~\cite{Jido}. 
The \siin\  resonance is known to decay  into the $N\eta$ 
channel with a large branching ratio of 30--55\%~\cite{PDG}, 
while other nucleon resonances in this energy region hardly decay to 
the $N\eta$ channel. 
Therefore, the excitation and decay of the \siin\  resonance 
is a dominant feature of $\eta$ photoproduction off the nucleon 
in the region of photon energies below 1000~MeV~\cite{Saghai,ETA-MAID}. 
It is, thus, expected that the properties of the \siin\  resonance 
in the nuclear medium can be studied through $\eta$ 
photoproduction off nuclei. 

 The measurements of the $A$($\gamma,\eta$) reactions 
have been reported by 
R\"{o}big-Landau  et al. 
on C, Ca, Nb and Pb for \Eg $<$ 800~MeV~\cite{Mainz_A}, 
and by Yorita et al. on C, Al and Cu 
for \Eg $<$ 1000~MeV~\cite{Yorita}.
In both measurements, the \siin\  resonance 
is clearly observed  
in the excitation function, 
which can be reproduced by a calculation 
taking into account the basic effects of the nuclear medium with
parameters 
deduced from the total cross section of the \gpep\  reaction. 
It seems, however, that the success of the interpretation of 
the $A$\gae reaction with the \siin\  resonance alone 
is partially due to the lack of the quality in the previous data 
for \Eg $>$ 800~MeV~\cite{Yorita} 
as well as those of the \gpep\  reaction.

In the last several years, there were essential progresses
in  experimental and theoretical work
on the \gpep\  reaction.
For the experimental side, precise measurements 
for \Eg $>$ 800~MeV have improved considerably 
the available data base~\cite{Mainz_p,Ajaka,Armstrong,
Thompson,Graal,CLAS,Elsa}. 
This led theoretical analyses to be more reliable for
including contributions of all the  resonances in this energy region 
as well as direct $\eta$ production  processes. 
Of particular interest is the fact that 
both of the analyses performed 
by Saghai et al.~\cite{Saghai} and 
by Chiang et al.~\cite{ETA-MAID}
have come to the same conclusion that another \sii\  resonance, 
\ssiin, also contributes in the total cross section 
of the \gpep\  reaction in such a way that the two \sii\  resonances 
interfere destructively. 

All these arguments raised the interest to study 
the behavior of the \siin\  resonance  again 
by measuring $A$\gae reactions with improved quality 
for the photon energies higher than 800~MeV. 
In this letter, we  present 
the experimental results and compare them 
with calculations based on the Quantum Molecular Dynamics model (QMD)
which is improved so as to include other processes
 than the \siin\  resonance.


 The experiment was performed at the Laboratory of Nuclear Science (LNS) 
in Tohoku University by using tagged photon beams from the 1.2 GeV
Stretcher-Booster Ring~\cite{STB}. 
Two series of measurements were carried out in different setups:
the first one at the photon beam line 1 in the experimental hall 2 and 
the second at the photon beam line 2 in the GeV-$\gamma$
experimental hall. 
The former tagging system is described 
in detail in Ref.~\cite{BM4tagger} 
and a part of data obtained in the first series was reported in 
Ref.~\cite{Yama2}. 
Photon beams of the same quality can be used at both 
beam lines. In the present work, the photon energy was covered 
from 600 to 850~MeV 
with $E_e$ = 920~MeV and from 800 to 1120~MeV with $E_e$ = 1200~MeV.
The total tagged photon intensity was about 10$^7$~Hz  with a duty factor
 of about 80\%. 
The size of the beam at the target position was about 6 mm (rms). 
The targets used were C and Cu with thicknesses of 40 and 5 mm, respectively.

Two photons from an $\eta$-meson were detected by an electromagnetic calorimeter 
consisting of 206 pure CsI crystals with plastic veto counters. 
The shape of the crystal is truncated-trapezoidal with a hexagonal cross
section and its thickness is 30 cm for 148 pieces (type-A) and 25 cm for
58 pieces (type-B); 
the performance of the type-B is described in detail in
Ref.~\cite{CsI}. 
In the first series of the measurements, 
they were assembled to 6 blocks and placed on three turn tables 
to change detector positions as reported in Ref.~\cite{Yama2}. 
In the second series, they were rearranged to 4 blocks placed 
in such a way that 
two forward blocks covered 
angles $15^\circ<\theta<72^\circ$ with respect to the beam direction 
and  angles $-17^\circ< \phi< +17^{\circ}$ with respect to 
the horizontal plane 
and two backward blocks $95^{\circ} < \theta < 125^{\circ}$ 
and $ -12^\circ< \phi <
 12^\circ$ for both sides of the beam direction. The different arrangements of 
crystals served to check the acceptance of the detection system.

All the data were collected 
using a similar data acquisition system as reported in Ref.~\cite{Yorita}.
In the present work, the main trigger for the data acquisition 
required at least one signal from the tagging counters and 
two signals from the CsI detectors. 
The maximum counting rate of a CsI detector was about 10~kHz
and that of a tagging counter was about 200~kHz.
The dead time of the data taking was  about 8\%.
A time resolution for $e$-$\gamma$ coincidences
of 800~ps  (FWHM) was achieved 
and the chance coincidence ratio was about 3\%.

\begin{figure}
\includegraphics[width=100mm]{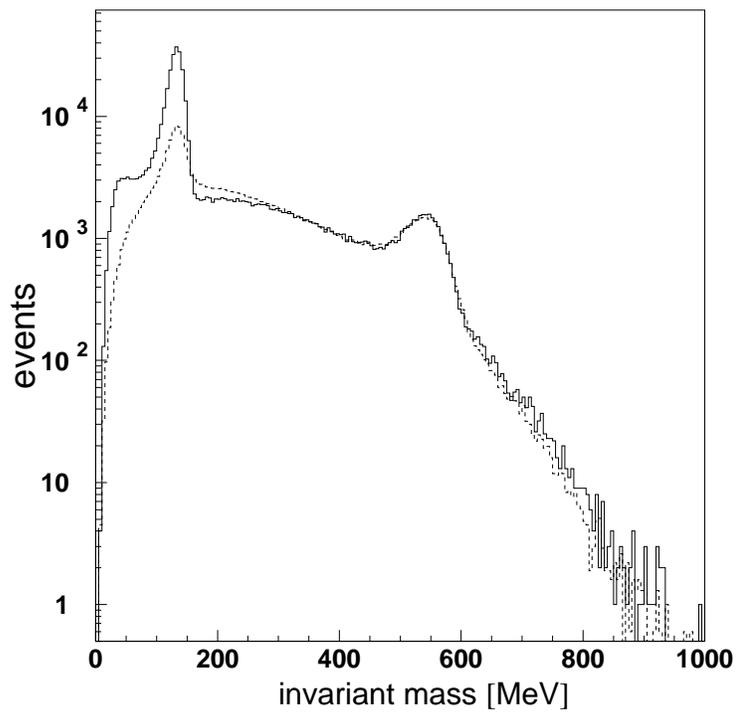}
\vspace*{0.5 cm}
\caption{
Invariant mass spectrum reconstructed from two photons.
The solid line represents the experimental data and 
the dotted line the result of the simulation.
}
\label{invmass}
\end{figure}

The $\eta$-mesons were identified via their two photon decay
with an invariant mass analysis.
In Fig.~\ref{invmass}, the invariant mass spectrum ($M_{\gamma\gamma}$) measured in the
present work is shown by the solid line. 
Two prominent peaks corresponding to
$\pi^0$ and $\eta$ mesons are clearly seen 
on the continuum background,
which is considered to originate mainly from multi $\pi^0$ events.
We simulated two $\pi^0$ production process 
by the Monte Carlo simulation.
The result is shown by the dotted line in Fig.~\ref{invmass}.
The shape is well fitted with an exponential function,
exp($aM_{\gamma\gamma}^2+bM_{\gamma\gamma}$). 
In order to deduce double differential cross sections, 
$d^2\sigma/d\theta/dp$,  
the invariant mass spectrum was constructed 
for the polar angle from 0$^\circ$ to 110$^\circ$ by 10$^\circ$ steps
and for the momentum from 0 to 1100~MeV by 100~MeV steps.
The yield of $\eta$-mesons in each spectrum was deduced by 
subtracting the background events in the $\eta$ mass region,
which were estimated 
with the function 
fitted to the continuum
for each bin of the incident photon energy and
the $\eta$-meson polar angle and momentum.
 Absolute cross sections were deduced  
by taking into the account the thickness of the targets, 
tagging counter counts, a tagging efficiency, 
a geometrical acceptance  and 
the branching ratio ($\eta\rightarrow\gamma\gamma$)~\cite{PDG}.
The tagging efficiency was measured with a total absorbing lead glass
detector positioned in the direct beam. 
The geometrical acceptance of the detection system was calculated
by the Monte Carlo simulation
based on GEANT3~\cite{geant3}.
The systematic uncertainties of the overall normalization come from
photon flux (1\%), 
background determination (5\%) 
and the geometrical acceptance  (5\%).
Consequently the overall systematic uncertainty is 7\%.

\begin{figure}
\begin{center}
\includegraphics[width=110mm]{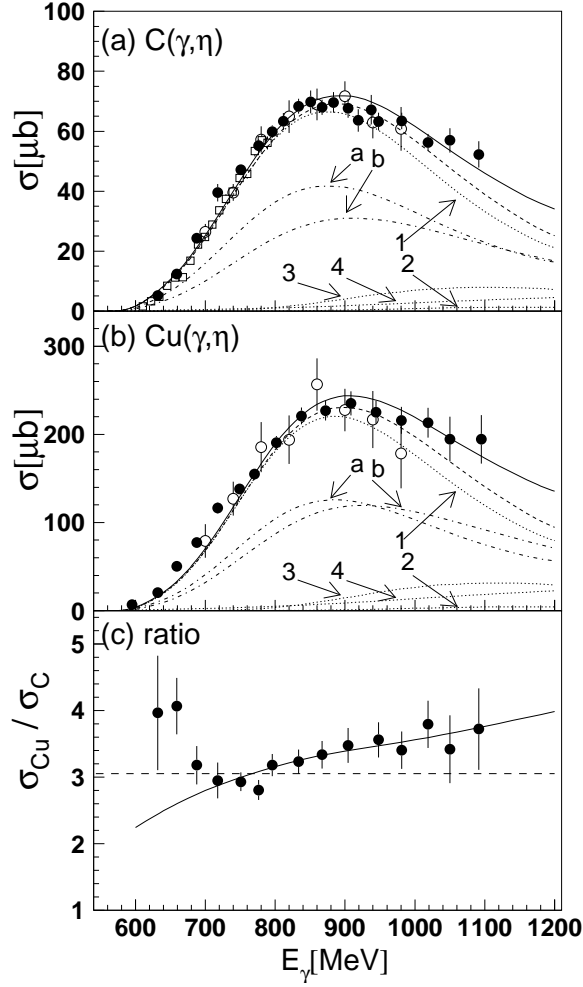}
\end{center}
\caption{Cross sections of $\eta$ photoproduction measured on C (a) and
 Cu (b). 
 The data measured in the present work are indicated by the solid circles, 
 those at KEK {\protect\cite{Yorita}}
 by the open circles and at Mainz {\protect\cite{Mainz_A}} by the squares. 
The solid line is the result of the present QMD calculation. 
Contributions of various processes are plotted by the dotted lines labeled with
 numbers; label 1 for the double \sii\  resonance, 
label 2 for the \pn,  
label 3 for the \dn,   
and label 4 for the direct processes. 
 The dot-dashed lines are contributions due to protons (label a) 
 and neutrons (label b). 
The calculation in the previous
 work by Yorita et al. {\protect\cite{Yorita}} is plotted by the dashed line. 
 (c) Ratio of the cross section of the  Cu\gae reaction 
 to that of the C\gae reaction. 
The dashed line shows the ratio of $A^{2/3}$ ($A$: mass number) 
and the solid line is the result of the QMD calculation.
}
\label{crs}
\end{figure}

Differential cross sections of the \gae reaction were deduced 
for the polar angles from 0$^\circ$ to 110$^\circ$ 
with respect to the photon beam direction
by integrating the double differential cross sections. 
 We show excitation functions of the $\eta$ photoproduction
 cross section,
which were deduced by integrating 
differential cross sections 
for 0$^\circ < \theta < 110^\circ$, 
on C and Cu targets 
in Fig.~\ref{crs}(a) and~\ref{crs}(b), respectively.
Missing yields for $\theta>110^\circ$ were estimated to be 2\% 
of the integrated values at most, and the total cross section in the
present work is the angle integrated one.
For comparisons,
also plotted are the previously reported data on C 
indicated with open squares up to 800~MeV~\cite{Mainz_A}
 and with open circles up to 1000~MeV~\cite{Yorita} 
and on Cu with open circles up to 1000~MeV~\cite{Yorita}. 
It can be said that the present data 
and the reported ones are in good agreement. 
Moreover, the statistical accuracy is much improved for the photon
energies higher than 800~MeV. 
The shape of the total cross section for C and Cu 
is quite similar as expected. 
The cross section increases rapidly from the threshold energy (561~MeV for C
and 550~MeV for Cu), shows a broad bump structure which has the maximum 
at around 850~MeV, and gradually decreases 
as the photon energy increases. This trend has been known from the previous
 investigations~\cite{Yorita} 
to be basically due to the excitation of the \siin\  resonance in a
 nucleus.
The present data for C and Cu may serve for detailed comparisons with model calculations. 

 In Fig. \ref{crs}(c), 
ratios of the cross section of Cu to that 
of C ($\sigma_{\rm Cu}$/$\sigma_{\rm C}$)
are plotted against the  photon energy. 
One can roughly say that the observed $\eta$-mesons are mainly emitted 
from the surface region of the nucleus and those emitted in deeper region are absorbed 
in the nuclear medium, since the ratios are close to 3.05 
(the dotted line), corresponding to the ratio of $A^{2/3}$ for Cu to C. 
However, 
there exist non-negligible and systematic deviations from the $A^{2/3}$ dependence for 
photon energies larger than 800~MeV; the ratio becomes about 3.5 at 
about 900~MeV. 
This requires more careful and detailed analysis.


 In order to explain the present data, we have performed 
a QMD model calculation 
in a different way from the previous one~\cite{Yorita} as follows. 
At first, the proton and the neutron are treated independently 
so as to see the effect of the difference of the 
elementary cross sections for \gpep\  and \gnen. 
This modification is necessary, because rather large difference 
between the total cross sections of \gpep\  and
\gnen\  has been predicted 
by the unitary isobar model, $\eta$-MAID~\cite{ETA-MAID}.
The simple relation 
$\sigma$(\gnen)/$\sigma$(\gpep) = 2/3,
established empirically for \Eg $<$ 800~MeV~\cite{Mainz_D} 
and used in the previous model calculations~\cite{Yorita,Mosel},
might not be correct at the higher energy region.
 Secondly, the effect of the interference between two \sii\ 
resonances in this energy region, 
\siin\  and \ssiin, is included in the calculation as the form of the cross 
section. 
The reason of this modification is as follows. 
Saghai et al.~\cite{Saghai} and Chiang~et al.~\cite{ETA-MAID} 
analyzed the total cross section of the \gpep\  reaction.
They came to the same conclusion 
that the experimentally observed cross section below 1100~MeV is not only 
due to the \siin\  resonance but also due to the \ssiin,
and both resonances make a destructive interference 
in the cross section of $\eta$ production off the nucleon. 
In the present QMD calculation, it is impossible to treat directly the
transition amplitudes, and, thus, 
the calculated cross section including a destructive interference 
term is incorporated instead of an incoherent sum of two single resonances
which cannot reproduce the experimental data well.

\begin{figure}
\begin{center}
\includegraphics[width=120mm]{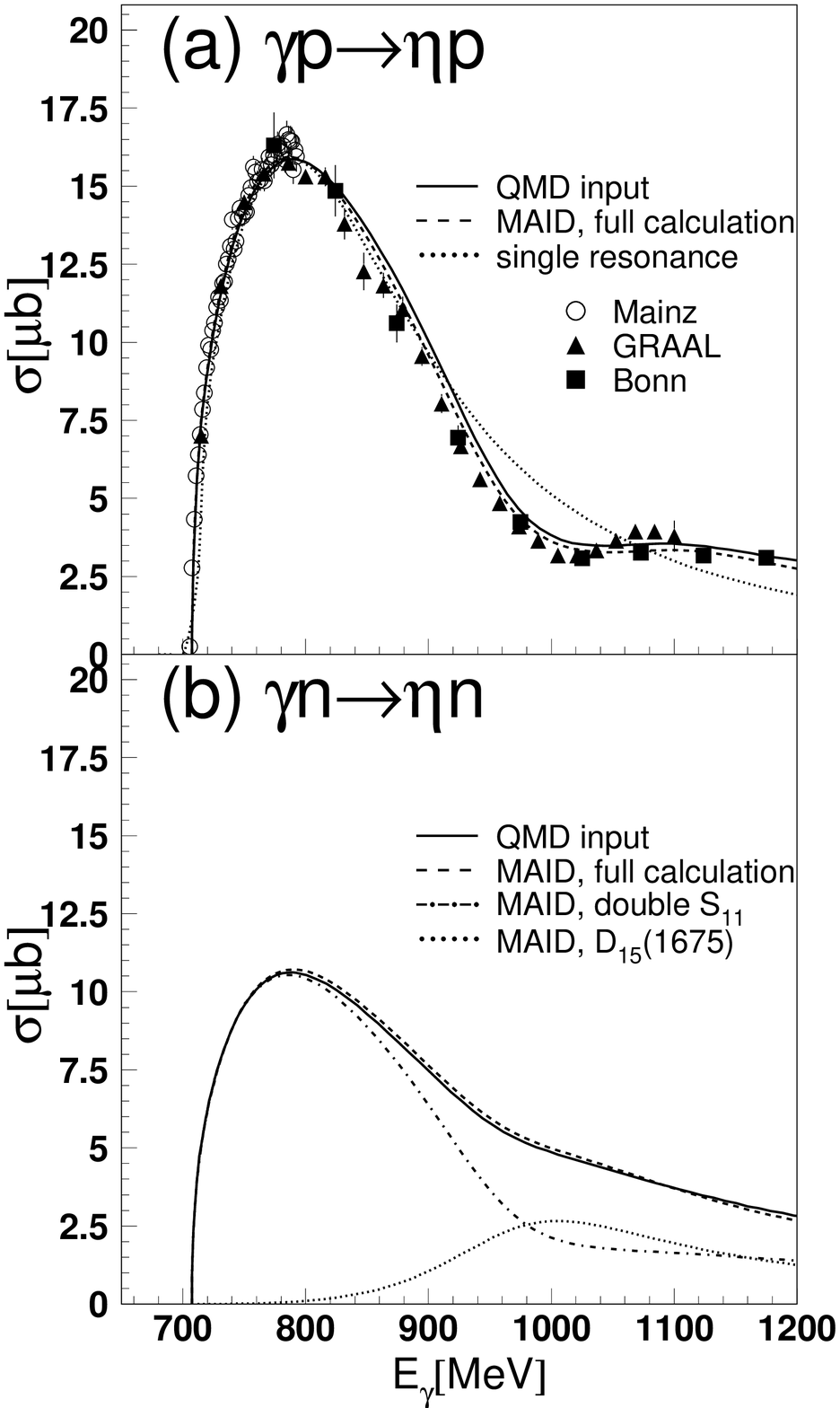}
\end{center}
\caption{
Excitation function of cross sections for the elementary reaction 
by calculation based  on the $\eta$-MAID: (a) the \gpep\  reaction and (b) the \gnen\  reaction. 
The solid lines are used in the present QMD calculation and the dashed lines
 are the results of the full $\eta$-MAID including all the resonances 
and the  direct process. 
In (a), the excitation function used in the previous work 
by Yorita et al. {\protect\cite{Yorita}} is also plotted by the dotted
 line as well as experimental data 
from Mainz {\protect\cite{Mainz_p}}, GRAAL {\protect\cite{Graal}} 
and Bonn {\protect\cite{Elsa}}. 
In (b), contributions of the two resonances, \siin\ and \ssiin, 
destructively interfered, are shown by the dot-dashed line, 
while those of the \dn\  are shown by the dotted line. 
}
\label{peta}
\end{figure}

 In Fig. \ref{peta}, the total cross sections of the $N$\gae reaction are
 shown in order to explain input quantities to the present QMD
 calculations. The experimental cross sections of 
the \gpep\  reaction are shown in Fig. \ref{peta}(a); 
data plotted with circles from Ref.~\cite{Mainz_p},
triangles from Ref.~\cite{Graal} and squares from Ref.~\cite{Elsa}.
Also shown are the results of the $\eta$-MAID calculation, 
on which we have based 
for the elementary cross section. The characteristic feature of the
experimental data is a broad peak followed by a flat region 
with a small dip at 1010~MeV. 
The single resonance excitation can reproduce 
only the broad resonance shape but fails 
to reproduce the dip and flat as indicated by the dotted line,
which corresponds to the elementary cross section used 
in our  previous analyses~\cite{Yorita}. 
The dashed line is the result of the full $\eta$-MAID calculation 
which includes all the resonances in this energy region 
with direct $\eta$ production processes. 
As mentioned above, the essential point is 
the destructive interference of the \siin\
and the \ssiin\  resonances, 
which reproduces the dip and flat behavior  very well. 
Another non-negligible process is the excitation of the \pn\  resonance 
which slightly contributes to the flat region around 1100~MeV. 
Therefore,
we have included three resonance excitations, 
\siin, \ssiin, and \pn,
and the direct processes
in the cross section of the \gpep\  reaction 
for the QMD calculation. 
The total cross section of the elementary 
\gpep\  reaction is calculated practically by summing up the cross sections of the 
double \sii, the $P_{11}$, and the direct processes, 
although the exact calculation should be 
the square of sum of the amplitude of each process. 
The solid line in Fig.~\ref{peta}(a) is the elementary \gpep\ 
cross section used in the present QMD calculation, being slightly larger 
than the full calculation of the $\eta$-MAID.   

 For the cross section of the \gnen\  reaction, no experimental data have 
been reported 
so far. 
We, again, follow the $\eta$-MAID
calculation as shown in Fig. \ref{peta}(b). 
The dashed line corresponds to the full calculation of 
the $\eta$-MAID including all the resonances and the direct processes. 
We select the double \sii\  resonance, the \dn\  resonance as major processes 
of excitations and 
the direct processes 
for the QMD input. 
The $\eta$-MAID calculation predicts rather large cross sections 
through the \dn\  resonance excitation, 
which is essentially prohibited in the \gpep\  reaction 
by the Moorhouse selection rule~\cite{Moorhouse}. 
The cross section through the two \sii\  resonances is calculated 
with the destructive interference term. 
It shows a dip at around 1000~MeV as shown by the
dot-dashed line in Fig. \ref{peta}(b),
where one sees that the large contribution from the \dn\  
resonance, which is plotted by the  dotted lines,  fills 
the dip and appears like a shoulder of the \siin\
resonance. 
The sum of the cross sections of the selected process, which is employed as the elementary 
\gnen\  reaction, is plotted by the solid line. 
The difference from the full calculation is very small as in the case
for the \gpep\  reaction. 

Having discussed the elementary cross section, 
we now return to the C\gae and Cu\gae reactions. 
 In Figs. \ref{crs}(a) and~\ref{crs}(b), the cross sections obtained 
in the present work are compared with the QMD 
calculations. 
As mentioned above, the elementary cross sections of the \gpep\  and 
\gnen\  reactions are treated independently; 
they are the solid lines in Fig.~\ref{peta}(a) and
Fig.~\ref{peta}(b) for proton and neutron, respectively. 
The $\eta$-emission probability  through the resonance excitation
 is calculated according to the Breit-Wigner resonance formula 
for the \pn\  and the \dn\   resonances.
For the \sii\  resonance excitation in the present work, 
however, the cross section including the interference of the two \sii\
resonances 
is used as if a resonance of the mixed state which is not described 
by the single Breit-Wigner formula is excited,
and the lifetime and the decay 
branch of the \siin\  resonance is applied to the mixed state. This approximation 
seems to be allowed, since the $\eta$-meson is mainly emitted via the
\siin\  resonance. 
The resonance parameters in the $\eta$-MAID are used for each resonance, and  are 
summarized in Table 1.

 It should be noticed for the QMD calculations 
in the present work that the effective 
energies of the incident channel are calculated 
for each photon-nucleon collision with a nucleon 
bound in a mean field potential. 
The effective energy which results in the reduction of the $\eta$
yield for the threshold region has been discussed~\cite{Mosel,Maru}
and the result of such calculation
improves the reproduction of the cross sections below 800~MeV. 
Thus, we have calculated the effective total  energy  followed as
$W=\sqrt{s}-U$,
where $W$ is the effective total energy, $\sqrt{s}$ is the
c.m. energy of the incident photon and a nucleon in the nucleus 
and $U$ is the nucleon potential 
calculated from the mean field potential in Ref.~\cite{Niita}.
As described in Ref.~\cite{Yorita}, the Fermi motion of nucleons, 
the Pauli blocking, collisions of nucleon resonances 
with nucleons and the absorption of the $\eta$-mesons 
are taken into account in the calculation.


In Fig. \ref{crs}(a) and~\ref{crs}(b), the dashed line is the results 
of the QMD calculation 
in which only the \siin\  resonance
 is incorporated with the assumption of 
$\sigma_n$/$\sigma_p$  = 2/3.
This corresponds to the previous calculation in Ref.~\cite{Yorita}.
For both  C\gae and Cu\gae reactions, the calculation reproduces data
up to 950~MeV. 
However, it underestimates the yield for \Eg $>$ 1000~MeV. 
The solid line corresponding to the new recipe covers the deficit 
and reproduces the data well up to 1100~MeV. 
Contributions of each process are also shown 
in Fig.~\ref{crs}(a) and~\ref{crs}(b). 
As can be seen, in addition to the largest contribution 
of the double \sii\  resonance
indicated by the dotted line 1, 
the contributions of the \dn\  resonance 
and the direct processes are expected for \Eg $>$ 900~MeV. 
The present calculation suggests that more than 18\%  of the cross
section at 1000~MeV  originates  from other processes 
than the excitation of the \sii\  resonance. 
It is of particular interest that the  \dn\  resonance
plays an important role for higher photon energies, 
since only neutrons can be excited 
in a naive quark model.
The dot-dashed lines labeled a and b in Fig. \ref{crs}
correspond to the contributions of protons and of neutrons,
respectively. 
The ratio of the contribution of neutrons to that of 
protons is nearly 0.67 for C and 0.84 for Cu at  \Eg $<$ 800~MeV,
where only the \sii\  resonance formation process can contribute, 
and becomes 0.97 for C and 1.23 for Cu 
at around \Eg $=$ 1100~MeV due to the existence of the \dn\  resonance.
The change of the contribution of neutrons to protons 
may explain the change of the ratio of the total cross section 
$\sigma_{\rm Cu}$/$\sigma_{\rm C}$
plotted in Fig.~\ref{crs}(c),
where the calculated ratio
is also shown by the solid line. 
The calculation explains the trend of the ratio
very well, although it fails for the lowest two points.  

The above discussion on Fig.~\ref{crs}
requires at least the following; 
the elementary cross section of the \gnen\  reaction  exceeds 
that of the \gpep\   
for \Eg $>$ 1000~MeV, 
due to another process besides the \sii\  resonance formation. 
Recently, Kuznetsov et al. reported the $n\eta$ and $p\eta$ coincidence
measurements in the D\gae reaction~\cite{Graal_D}. 
Their result that yields of the $n\eta$ coincidence events 
are larger than those of the $p\eta$ events at around 1000~MeV 
is consistent with the present interpretation.

\begin{figure}
\begin{center}
\includegraphics[width=120mm]{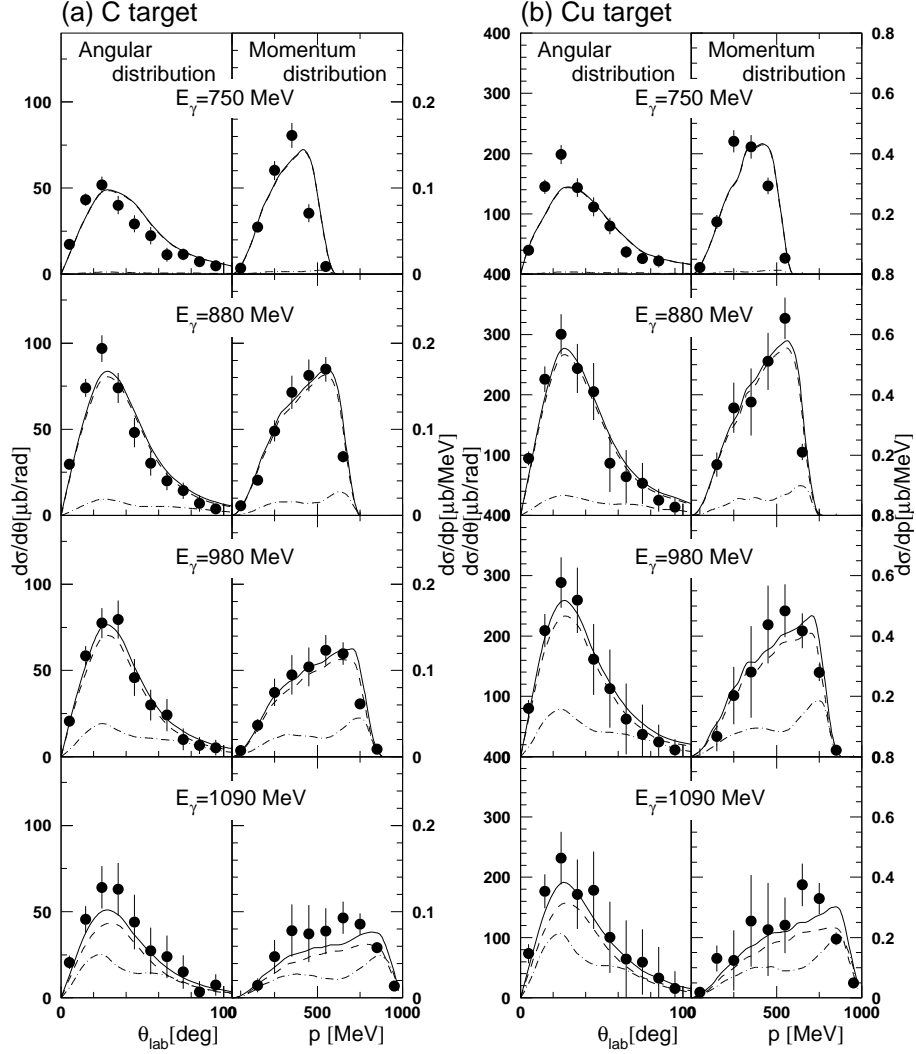}

\end{center}
\caption{Angular distributions, 
$d\sigma$/$d\theta$,   
and momentum distributions, 
$d\sigma/dp$,
of the ($\gamma,\eta$) reaction measured on C (a) and Cu (b).
The present results are plotted with the closed circles
for \Eg = 750, 880, 1000, and 1100~MeV.
The solid line is the results of the QMD calculation and 
the dashed one represents without the contribution of \dn,
and the dot-dashed line shows only the contribution of \dn\
resonance multiplied by 4. 
}
\label{momdegdis}
\end{figure}

 Additional effects that might possibly give rise deviations from the
pure \sii\  resonance formation may be seen in the angular and momentum
distributions of the emitted $\eta$-mesons. 
They are shown in Fig.~\ref{momdegdis} for \Eg = 750, 880, 980, and 1090~MeV, and compared 
with the QMD calculations. 
The results for the C target are shown in Fig. \ref{momdegdis}(a), 
and the Cu target in Fig. \ref{momdegdis}(b). 
All the angular distribution data show a broad structure peaked 
at around 30$^\circ$. This is a characteristic of the quasi-free $s$-wave $\eta$ 
production. The solid lines in the figure are the results of the QMD
calculation, and the experimental data for both angular and momentum
distributions are essentially well reproduced by the corresponding
calculations. The dashed lines shows the results without the contribution
of the \dn\  resonance and the dot-dashed lines 
correspond to the contribution of the \dn\  resonance multiplied by 4. 
As shown, the contribution of the \dn\  resonance has a different structure
 in these distribution because of the $d$-wave $\eta$ emission.
Since the contribution of the \dn\  resonance is not large, 
the present data, unfortunately, cannot give a firm evidence for the excitation of 
the \dn\  resonance in both reactions.

 As we have discussed above, the comparison of new data with the QMD model 
calculations suggests strongly that the cross sections of $\eta$
photoproduction off nuclei contain non-negligible quantities through the
process other than the \siin\  resonance formation for the photon energies
above 850~MeV. 
Thus, in order to investigate the change of the \sii\  properties 
in the nuclear medium 
such as the mass shift proposed in Ref.~\cite{Jido}, 
one needs precise data of the cross section for the \gnen\  reaction. 
The present work has shown that the use of the elementary cross sections
of $\eta$-MAID can reproduce the experimental data very well but remains an
interesting subject for future investigations.

In summary, the $\eta$ photoproduction cross sections were measured 
on the C and Cu targets for the photon energies between 600 and 1100~MeV. 
The excitation functions of the total cross section as well as 
angular and momentum distributions were in 
quantitative agreement with the QMD model calculations in which the cross sections 
proposed in the $\eta$-MAID model were used for the elementary reactions
\gpep\  and \gnen. 
The agreement suggests that there is a difference in the shape of the cross 
sections between proton and neutron in a nucleus. 
In order to discuss the change of the properties of the \siin\  resonance, 
the cross section of the \gnen\  reaction experimentally measured
is highly desirable to be incorporated
in the model calculations.

\ack
 We would like to thank the accelerator staff at LNS for their support 
during the experiment. 
This work was partly supported by the Grant-in-Aid for Scientific Research 
(Nos. 1040067, 07740197, and 15340069) of the Ministry of Education of Japan.

\clearpage

\begin{table}
\caption{\label{tab:table1}Parameters of nucleon resonance used in
 our calculation. 
$A^{p,n}_{1/2,3/2}$ are photoexcitation helicity amplitude of nucleon 
resonances and $\beta_{\eta N}$ are $N\eta$ decay branching ratio.
Those of the last columm are 
used in the previous analysis \cite{Yorita}.}
\begin{tabular}{cccccccc}\hline
N$^*$ & Mass  &Width & $\beta_{\eta N}$ & $A^p_{1/2}$ & $A^p_{3/2}$  & $A^n_{1/2}$  & $A^n_{3/2}$  \\ 
    &  [MeV]  & [MeV] & [\%] & \multicolumn{4}{c}{[10$^{-3}$GeV$^{-1/2}$]}
   \\ \hline
\siin  & 1541 & 191 & 50 & +118 & ---  & $-$96  & ---  \\ 
\ssiin  & 1638 & 114 & 7.9 & +68 & --- & $-$56  & ---  \\ 
\pn & 1721 & 100 & 26 & +23 & --- & 0  & ---  \\ 
\dn & 1665 & 150 & 17 &  0  & 0 & $-$43  & $-$58  \\ \hline
\siin  & 1542 & 150 & 55 & +102 & --- & $-$83  & ---  \\ \hline
\end{tabular}
\end{table}

\clearpage

\clearpage

\clearpage

\end{document}